\documentclass[aps,prl,twocolumn,preprintnumbers,amsmath,amssymb,superscriptaddress,floatfix]{revtex4}
\usepackage[dvips]{graphicx}
\usepackage{mathptmx}

\begin{document}

\title{Magneto-electric coupling in zigzag graphene nanoribbons}
\author{J. Jung} \email{jeil@physics.utexas.edu}
\affiliation{Department of Physics, University of Texas at Austin, USA}
\author{A. H. MacDonald}
\affiliation{Department of Physics, University of Texas at Austin, USA}

\begin{abstract}
Zigzag graphene nanoribbons can have magnetic ground states 
with ferromagnetic, antiferromagnetic, or canted 
configurations, depending on carrier density.  
We show that an electric field directed across the 
ribbon alters the magnetic state, favoring 
antiferromagnetic configurations.  This property can be 
used to prepare ribbons with a prescribed spin-orientation 
on a given edge.
\end{abstract}
\pacs{73.63.-b, 71.15.Mb, 73.40.Jn, 05.60.Gg}
\maketitle

\noindent
{\em Introduction---}
Expanding techniques that can achieve electrical control of spin
is a key goal of both metal and 
semiconductor spintronics.\cite{spintronics,trends_physics}
In metal spintronics, for example, electrical spin-transfer torque\cite{stt}
research seeks to amplify the potential\cite{mram} of technologies  based on 
giant magnetoresistance \cite{gmr} and   
tunnel magnetoresistance. \cite{tmr}
The aim of research on dilute magnetic semiconductors\cite{dms}
is to create semiconductor materials in which magnetic properties
are as sensitive to doping and external gate potentials as electrical properties.  
Recent interest in the spin Hall effect \cite{spinhall} 
and the topological magnetoelectric effect \cite{topological_insulators} 
is motivated by a search for effects which enable
electrical control of spin in non-magnetic materials. 
In this context it is interesting to address the 
possibility of interesting magnetoelectric effects in
graphitic material.  
The physics of zigzag graphene ribbons and edge terminations
has received considerable attention recently
\cite{fujita,hikihara,dutta,leehosik,sasaki,son_gap,son_half,
pisani,yazyev,joaquin,pohang,superexchange,rhim,bhowmick,islands, ciraci,
nakada,waka,ezawa,brey,yaowang,white,sandler,han,chem_ribbon,etching,
jarillo,rice,dresselhaus,zettl,chuvilin}.
Magnetism is expected in any graphitic material containing ribbon segments with 
zigzag edge \cite{fujita,hikihara,dutta,leehosik,sasaki,son_gap,son_half,
pisani,yazyev,joaquin,pohang,superexchange,rhim,bhowmick,islands, ciraci}
terminations, for example highly defected bulk graphitic material. \cite{defects,esquinazi,yazyev1,maggraphite} 
Thanks to progress in structural control
of graphene flakes and related materials \cite{graphenereviews},
prospects for mastering graphitic magnetism have improved.
The perfect zigzag nano-ribbon studied in this 
Letter may be viewed as a model system in which graphitic 
magnetism is exhibited in its simplest and most essential form. 

The magnetic ground state of a zigzag nanoribbon has  
collective moments localized near its edges.  In the absence of a transverse 
electric field, a doped nanoribbon has either full or partial 
orientational alignment\cite{noncollinear, doping} 
between moments on opposite edges, as illustrated schematically in Fig.[1].  We show that
an external electric field applied across the ribbon can control
the relative orientation angle $\theta$, and that 
this property can be used to prepare ribbons with a prescribed spin-orientation 
on a given edge.   

\begin{figure}[tbp]
\label{ex_distr}
\begin{center}
\includegraphics[width=6.5cm,angle=90]{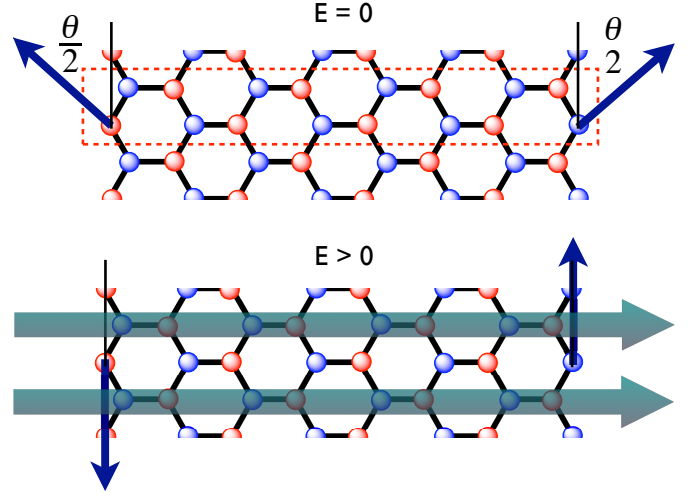}
\caption{
Schematic representation of the influence of a lateral electric field on 
zigzag ribbon magnetism. 
{\em Upper panel.}
Non-collinear ground state of a doped zigzag ribbon illustrating 
the angle $\theta$ between moment orientations on opposite edges. 
The red box indicates the unit cell of the ribbon.
{\em Lower panel.}
A lateral electric field drives the angle $\theta$ to $\pi$, an
antiferromagnetic configuration similar to that of an undoped ribbon. 
}
\end{center}
\end{figure}

\noindent
{\em Hubbard Model Mean-Field Theory---}
Since ab initio density-functional and simpler model Hamiltonian approaches 
make essentially identical predictions \cite{joaquin, bhowmick, islands, yazyev1},
we base our analysis of zigzag ribbon magnetism on a Hubbard model which allows the underlying physics to be 
identified more clearly.  The success of the Hubbard model has been shown 
to be due to the essentially local character of edge magnetism in graphene ribbons \cite{nonlocal}.
The Hubbard model mean-field Hamiltonian, 
\begin{eqnarray}
\label{eq:hmf}
H_{\sigma} &=& - \gamma_0 \sum_{\left< i,j \right>}  c^{\dagger}_{i \sigma}c_{j \sigma} +  \sum_{i}   c^{\dagger}_{i \sigma} c_{i \sigma} (v_{ext} + e {\cal E} y_i)  \nonumber \\
&+& \frac{U}{2} \sum_{i} \left[ \langle c^{\dagger}_{i \sigma} c_{i \sigma}\rangle   c^{\dagger}_{is} c_{is} 
- \langle c^{\dagger}_{i \sigma'} {\vec \tau}_{\sigma',\sigma} c_{i \sigma}\rangle  \cdot  c^{\dagger}_{i s'} 
{\vec \tau}_{s',s} c_{i s} \right], 
\end{eqnarray}
has a term which represents hopping between nearest neighbor $\pi$-orbitals with 
amplitude $\gamma_{0} = 2.6 eV$,
an external potential term which accounts for the transverse electric field,
and a mean-field interaction term.  
The operator $c^{\dag}_{i \sigma}$ creates a $\pi$ orbital electron at site $i$ with spin $\sigma$,
${\cal E}$ is the transversal electric field, $y_i$ is the position of lattice site $i$ along the ribbon 
width and ${\vec \tau}_{\sigma',\sigma}$ represents the elements of the three Pauli matrixes. 
As a convenience we 
include a constant term $v_{ext} = -U$ in the 
external potential which removes the interaction with a unit charge on 
each site from the mean-field quasiparticle energy. 
(All spin indices in Eq.~[\ref{eq:hmf}] are summed over.)  
Note that the mean-field interaction energy of an electron on site $i$ 
is spin-dependent and proportional to the density of opposite spin electrons.
Following Yazyev {\em et al.},\cite{yazyev1} we choose $U=3 eV$
a value slightly larger than estimates based on the local density approximation
used in some previous Hubbard model analyses \cite{joaquin, superexchange,  doping}.
We have used 1200 $k$-points in the Brillouin zone for the self-consistent calculations.
Zig-zag edge magnetism is very sensitive\cite{doping} to the net charge density of the ribbon $\delta n$,
which we refer to as doping whether due to chemical dopants or gate voltages 
and measure per repeat distance $a = 2.46 \AA$ along the edge.
The corresponding areal density is $\delta n_{2D} = \delta n / W$ where the ribbon width 
$W = \sqrt{3} N a/2$ and $N$ is the number of atom pairs 
per ribbon unit cell.

\begin{figure}[htb]
\begin{center}
\includegraphics[width=6.5cm,angle=90]{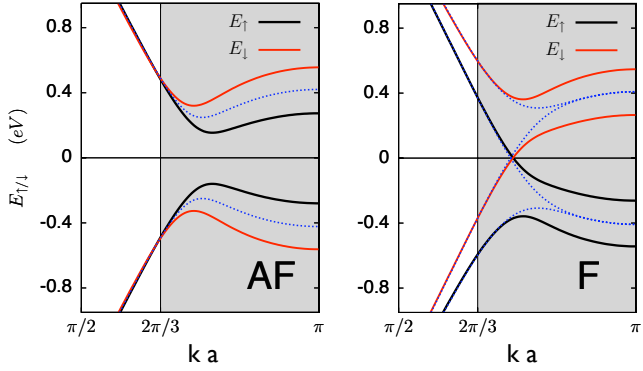}
\caption{
Edge state bands of a zigzag ribbon with $N=20$,  
nearest-neighbor hopping $\gamma_0 = 2.6 eV$ and on site repulsion $U=3 eV$
under a transverse electric field of ${\cal E} = 0.2 eV/nm$.
{\em Left panel}: 
In the AF case high-spin (see text) bands 
shift toward the Fermi energy and the low-spin bands move away.  
The wavefunctions of the occupied bands shift density to the low-energy edge. 
{\em Right panel}: 
In the F case $\Delta_{\cal E}$ shifts the wavevector at which majority and 
minority spin bands cross to slightly larger values of $|k|$.
The dotted lines show the bands at $\Delta_{\cal E} = 0$.    
}
\label{fig:hubbardmf}
\end{center}
\end{figure}

\noindent
{\em Edge-Only Model---}
The influence of a transverse electric field on 
electronic structure in neutral ribbons has been studied  
previously for both armchair \cite{armch_ef}
and zigzag cases \cite{son_half, rudberg, others_ef} using density functional
theory.
The essentials of zigzag-ribbon magnetism and of the transverse-field 
magnetoelectric effect are captured by an 
{\em edge-state-only} model\cite{joaquin,superexchange}; the qualitative discussion
below refers mainly to this model and to the special case of collinear magnetic states ($\theta=0$ or $\theta=\pi$),
but the numerical calculations and the phase-diagram results are 
based on solutions of the the full non-collinear $\pi$-band Hubbard model self-consistent field 
equations.  In the edge-only model the spin-dependent mean-field Hamiltonian
for collinear states  
takes the form\cite{superexchange,nonlocal}
\begin{equation}
\label{h3term}
H_{\sigma}\left( k \right) = (\sigma \Delta_{z}\left( k \right)  
+ \Delta_{\cal E} \left( k \right) ) \tau_{z}+ (\Delta_{0} \left( k \right) + h_{z} ) \sigma I 
+ t \left( k \right) \tau_{x}.
\end{equation}
Here the $\tau_{\alpha}$ are Pauli matrices which act on the {\em which edge} degree of 
freedom.  The terms proportional to $\tau_{z}$ in Eq.~[\ref{h3term}] therefore 
represent the difference in energy between left ($\tau_{z} \to 1$) and right 
($\tau_{z} \to -1$) edges for 
electrons of spin $\sigma$, whereas the term proportional to $\tau_{x}$ 
represents the momentum-dependent inter-edge hopping amplitude.  Zig-zag edge magnetism 
follows from the property\cite{superexchange} that
$t(k)$ vanishes rapidly with ribbon width in the part of the 
Brillouin-zone ($2\pi/3a < |k| < \pi/a$) in which edge states reside.
In Eq.~(\ref{h3term}), $\Delta_{z}\left( k \right)$ captures the difference between exchange energies on 
opposite edges, which vanishes in the $\theta=0$ (F) state in the absence of 
a transverse field, whereas $\Delta_{0}\left( k \right)$ captures the spin-dependence of 
the edge average, which vanishes in the $\theta=\pi$ (AF) state.
(An irrelevant spin and edge independent exchange energy has been dropped from $H_{\sigma}$.) 
Both exchange energies are large only for the edge-states ($|k|>2\pi/3a$ ).
$h_{z}$ accounts for Zeeman 
coupling to the ribbon spins by an external magnetic field when present.

\begin{figure}[tbp]
\label{ex_distr}
\begin{center}
\includegraphics[width=8cm]{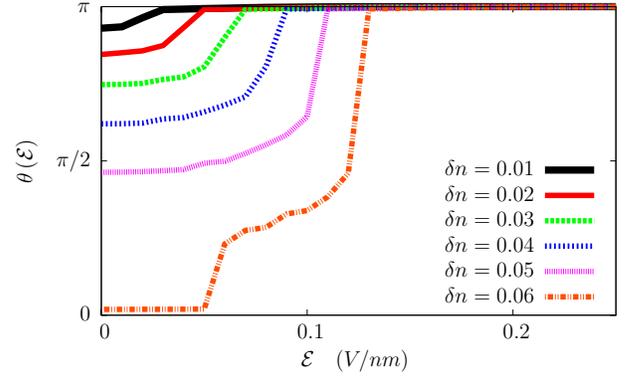} 
\caption{
Inter-edge relative spin orientation angle $\theta$ evaluated on the 
outermost edge atoms as a function of transverse electric field $\Delta_{\cal E}$
for a series of doping $\delta n$ values.
For $\Delta_{\cal E}=0$ doping leads immediately to a canting angle 
$\theta < \pi$ and eventually for $\delta_{n}$ larger than $\sim 0.05$ to a ferromagnetic state with 
$\theta = 0$.  A finite $\Delta_{\cal E}$ favors the AF state as explained in the text and drives $\theta$ 
toward its undoped value.
}
\end{center}
\end{figure}

\noindent
{\em Magnetoelectric Coupling in Undoped Ribbons---}
The eigenenergies of this Hamiltonian are 
\begin{eqnarray}
E_{\sigma}^{\pm} \left( k \right) = \sigma  (\Delta_{0} \left( k \right) + h_{z})  
 \pm  \sqrt{   \left(  \sigma \Delta_{z} \left( k \right)  + \Delta_{\cal E} \left( k \right)  \right)^2  + t^{2} \left( k \right)   }.
\label{efield}
\end{eqnarray}
Note that there are always four distinct eigenvalues in the ferromagnetic case, whereas the 
antiferromagnetic state bands occur in doubly-degenerate pairs when $\Delta_{\cal E} \to 0$.
For undoped ribbons the lowest two edge states bands are normally fully occupied.
In Fig.2 we plot ribbon band structures for both AF and F 
states of a neutral ribbon calculated 
using a constant transverse electric field of ${\cal E} = 0.2 \,\, V/nm$.
States that are shifted down (up) in energy relative to the $\Delta_{\cal E} = 0$ case
are localized on the low (high) potential side of the ribbon.
(Note that a constant field generates a $k$-dependent $\Delta_{\cal E}$ because of 
the $k$-dependence of the degree of edge state localization.)
In the F-state the Fermi energy is pinned to a 
band-crossing near $|k| = 2\pi/3a$ between the higher energy majority spin band and the 
lower energy minority spin band.  In the AF case we refer to the spin-orientation which dominates 
occupied states on the low (high) potential side of the ribbon as the low-spin (high-spin).
(In Fig.2, $\downarrow$ is the low-spin.)   A transverse spin shifts the energies of both 
occupied and unoccupied high-spins toward the Fermi level, lowering the gap.  
A sufficiently large transverse field will close the indirect gap, creating a half-metallic band
structure with only high-spin bands crossing the Fermi level.  This is the magnetoelectric
effect discussed in earlier\cite{son_half,others_ef,rhim} work.
The energy difference between F and AF states\cite{superexchange} is 
relatively unchanged by a transverse field.  Below we show that in doped ribbons a transverse field 
tilts the competition between F and AF states in favor of the latter, yielding a 
distinct and stronger magnetoelectric effect.

\noindent
{\em Magneto-electric coupling at finite doping---}
\begin{figure}[tbp]
\label{ex_distr}
\begin{center}
\includegraphics[width=8cm]{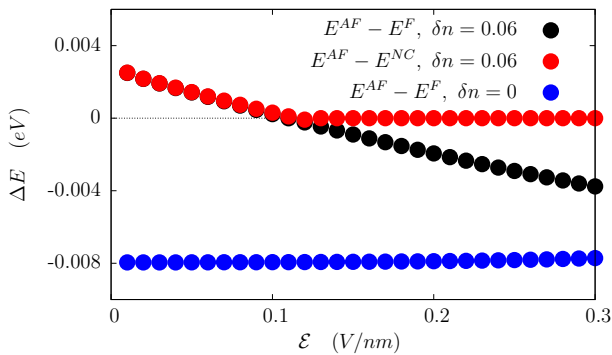}
\caption{
Energy differences (per ribbon unit cell)
between the spin collinear AF, F solutions and, for 
finite-doping, the minimum energy non-collinear (NC) solution.
The energy difference between AF and F solutions for $\delta n = 0$
has a weak transverse electric field dependence.
}
\label{figex}
\end{center}
\end{figure}
We consider for definiteness the case of n-type ribbons in which  
carriers are added by gate doping. 
In the absence of a transverse electric field, doping favors\cite{noncollinear,doping} the gapless 
F state over the gapful AF state.  The transition between undoped AF 
and doped F states occurs continuously by varying the relative orientation angle 
$\theta$ between its AF ($\theta=\pi$) and F ($\theta=0$) end points.   
The effect we discuss in this paper is based on the following simple observation
concerning the edge-state bands plotted in Fig. 2.
In the AF case the conduction band states which are occupied upon doping are 
high-spin antibonding states, which are localized on the low-energy side of the 
ribbon.  For the F state, on the other hand, there are occupied states in two 
bands, one localized on the high-energy side and one localized on the low-energy side.
The net effect is that a transverse field favors the AF state in doped zigzag ribbons.
In Fig.~3 we plot the relative angle between spin polarizations on opposite 
edges {\em vs.} lateral electric field for a series of different doping values.
These results were obtained by non-collinear spin self-consistent field calculations and 
confirm the expected magnetoelectric effect.  In Fig.4 we compare the 
transverse field dependence of the energy difference between F and AF states 
for doped and undoped systems.  The electric field strength 
required to convert F states into AF states in doped ribbons is 
much smaller than the field required to close the AF-state gap in the undoped case. 

\noindent
{\em Discussion---}
Typical results for the edge-state bands of both electron and hole doped 
zigzag ribbons with a transverse field strong enough to induce the $\theta=\pi$ state are 
illustrated in Fig.5.  Because of the partial occupation of the 
highest unoccupied band of the $n$-doped case and the lowest 
unoccupied band in the $p$-doped case, the $\theta=\pi$ state 
has an overall spin-polarization proportional in magnitude to the doping.
Note that these $\theta=\pi$ states are always half-metallic.
This ferromagnetic component of the order allows Zeeman coupling from an 
external magnetic field to fix the spin-orientation on each edge.
It follows from Fig.5 that the majority spins are high-energy spins in the 
$n$-doped case and low-energy spins in the $p$-doped case. 
For a fixed magnetic field direction, the spin-orientations on both edges 
can therefore be switched with a gate voltage, which changes the 
sign of the carrier density.  This remarkable property of zigzag edge 
magnetism has no parallel of which we are aware in any other magnetic system.    

\begin{figure}[tbp]
\begin{center}
\includegraphics[width=8cm]{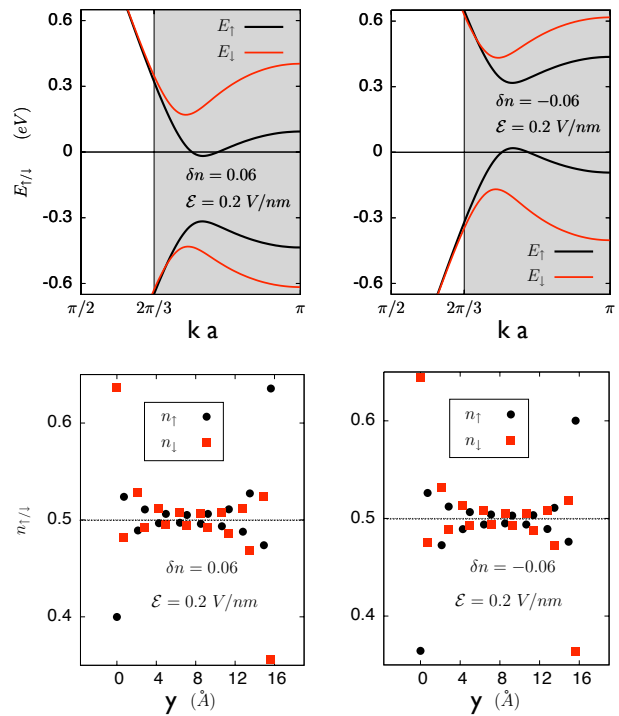} 
\caption{
{\em Top panel:} Band structure of $n$ and $p$ gate-doped $N=8$ zigzag graphene nanoribbons
in the presence of a transverse electric field, 
${\cal E} = 0.2 \,\, V/nm$, strong enough to produce a collinear $\theta=\pi$ ground state.  Note that the 
edge-state bands are half-metallic in both cases. 
{\em Bottom panel:}
Spin resolved occupation $n_{\uparrow / \downarrow} \left( y \right)$
across the robbon:
majority spin electrons ($\sigma = \uparrow$ for $n$-doped and $\sigma = \downarrow$ for $p$-doped)
in this figure accumulate on the high potential edge of the ribbon for $\delta n= 0.06$
and on the low-potential side for $\delta n= -0.06$. 
}
\label{fig:afef}
\end{center}
\end{figure}

In closing we remark that the one-dimensional character of zigzag edge
magnetism works against robust collective spin properties.
As analyzed in more detail elsewhere\cite{yazyev,rhim}, the 
consequences of reduced dimensionality are somewhat mitigated by the 
the substantial stiffness\cite{yazyev,rhim} of zigzag edge moments.
Nevertheless, robust magnetism in graphitic nanostructures will likely 
require exchange-coupled two-dimensional ribbon networks.  
The rather unique properties of graphitic magnetism 
discussed in this Letter motivate an effort to realize structures of 
this type.

{\em Acknowledgement.}
We acknowledge financial support from the Welch Foundation, NRI-SWAN, the DOE,
the Spanish Ministry of Education through the MEC-Fulbright program and 
Fan Zhang for his help with figures.


\begin{thebibliography}{99}

\bibitem{spintronics}
S. A. Wolf et al.
Science 294, 1488 (2001).
I. Zutic et al. 
Rev. Mod. Phys. 76, 323 (2004).
T. Dietl et al. "Spintronics", Elsevier (2008).

\bibitem{trends_physics}
D. Awschalom and N. Samarth,
Physics {\bf 2}, 50 (2009).

\bibitem{stt}
D. C. Ralph, M. D. Stiles
J. Magnetism and Magnetic Materials 320, 1190 (2008).

\bibitem{mram}
J. Akerman, 
Science 308, 508 (2005).

\bibitem{gmr}
P. Gr\"unberg, 
Rev. Mod. Phys. 80, 1531 (2008).

\bibitem{tmr}
M. Julliere,
Phys. Lett. 54A, 225 (1975).
J. Mathon and A. Umerski,
Phys. Rev. B 63, 220403 (2001).


\bibitem{dms}
H. Munekata, et al.  
Phys. Rev. Lett. 63, 1849 (1989).
H. Ohno et al.  
Nature 408, 944Ð946 (2000).
A. H. MacDonald et al.
Nature Materials 4, 195-202 (2005).
T. Jungwirth et al,
Rev. Mod. Phys. 78, 809 (2006).
 
\bibitem{spinhall}
J.E. Hirsch,
Phys. Rev. Lett. 83: 1834 (1999).
Sinova et al. 
Phys. Rev. Lett. 92, 126603 (2004).
S. Murakami et al., 
Science 301, 1348 (2003).

\bibitem{topological_insulators}
X.-L. Qi et al.
Phys. Rev B. 78, 195424 (2008).


\bibitem{fujita} 
                M. Fujita, K. Wakabayashi, K. Nakada, K. Kusakabe,
                J. Phys. Soc. Jpn. {\bf 65}, 1920 (1996).

\bibitem{hikihara}  
T. Hikihara, X. Hu, H.-H. Lin, and C.-Y. Mou
Phys. Rev. B {\bf 68}, 035432 (2003).

\bibitem{dutta} 
S. Dutta, S. Lakshmi, and S. K. Pati,
Phys. Rev. B {\bf 77}, 073412 (2008).

\bibitem{leehosik} 
H. Lee, Y.-W. Son, N. Park, S. Han, and J. Yu,
Phys. Rev. B {\bf 72}, 174431 (2005).
\bibitem{sasaki} 
K.-I. Sasaki, S. Murakami, R. Saito, 
               J. Phys. Soc. Jpn. {\bf 75}, 074713 (2006).
\bibitem{son_gap} 
Y.-W. Son, Marvin L. Cohen, and Steven G. Louie,
               Phys. Rev. Lett. 97, 216803 (2006).
\bibitem{son_half} 
Y.-W. Son, Marvin L. Cohen, and Steven G. Louie,
               Nature {\bf 444}, 347 (2006).            
\bibitem{pisani} 
L. Pisani, J. A. Chan, B. Montanari, and N. M. Harrison
Phys. Rev. B {\bf 75}, 064418 (2007).
\bibitem{yazyev} 
O. V. Yazyev, M. I. Katsnelson
Phys. Rev. Lett. {\bf 100}, 047209 (2008).

\bibitem{joaquin} 
J. Fern\'{a}ndez-Rossier,
Phys. Rev. B {\bf 77}, 075430 (2008).                 

\bibitem{pohang}
W. Y. Kim and  K. S. Kim,
Nature Nanotechnology 3, 408  (2008). 

\bibitem{superexchange}
J. Jung. T. Pereg-Barnea, A. H. MacDonald, Phys. Rev. Lett. 102, 227205 (2009). 

\bibitem{rhim}
J.-W. Rhim and K. Moon,
Phys. Rev. B {\bf 80}, 155441 (2009).


\bibitem{bhowmick}
S. Bhowmick and V. B. Shenoy,
J. Chem. Phys. 128, 244717 (2008).

\bibitem{islands}
J. Fern\'{a}ndez-Rossier, J. J. Palacios
Phys. Rev. Lett. 99, 177204 (2007).

\bibitem{ciraci}
M. Topsakal, H. Sevincli, S. Ciraci, 
App. Phys. Lett. 92, 173118 (2008) 

\bibitem{yaowang}
W. Yao, S. Yang, Q. Niu,
Phys. Rev. Lett. 102, 1 (2009).

\bibitem{nakada} 
               K. Nakada, M. Fujita, G. Dresselhaus, M. S. Dresselhaus,
               Phys. Rev. B {\bf 54}, 17954 (1996).

\bibitem{waka} 
             K. Wakabayashi, M. Fujita, H. Ajiki, M. Sigrist,
             Phys. Rev. B {\bf 59}, 8271 (1999).

\bibitem{ezawa} 
M. Ezawa,
              Phys. Rev. B {\bf 73}, 045432 (2006).
              
\bibitem{brey} 
L. Brey and H. A. Fertig, 
              Phys. Rev. B {\bf 73}, 235411 (2006).
              

\bibitem{white}
D. Gunlycke, D. A. Areshkin, L. Junwen, J. W. Mintmire, C. T. White,
Nano letters. 7, 3608 (2007);  
D. Gunlycke, H. M. Lawler, C. T. White, 
Phys. Rev. B. 75, 29-33 (2007).

\bibitem{sandler}
M. Zarea, C. Busser and N. Sandler, Phys. Rev. Lett. 101, 196804 (2008).

\bibitem{han}
M. Y. Han, Barbaros \"Ozyilmaz, Y. Zhang, and P. Kim, Phys. Rev. Lett. 98, 206805 (2007).

\bibitem{chem_ribbon} 
X. Li, X. Wang, L. Zhang, S. Lee, H. Dai,
Science {\bf 319}, 1229 (2008).

\bibitem{etching}
S. S. Datta, D. R. Strachan, S. M. Khamis and A. T. C. Johnson,
Nano Lett. {\bf 8} 1912 (2008).

\bibitem{jarillo}
L. C. Campos, V. R. Manfrinato, J. D. Sanchez-Yamagishi, J. Kong, P. Jarillo-Herrero,
Nano letters. 9, 2600 (2009)

\bibitem{rice}
L. Ci, L. Song, D. Jariwala, et al. 
Adv. Mat. 21, 4487 (2009)

\bibitem{dresselhaus}
X. Jia, M. Hofmann, V. Meunier, et al. 
Science 323, 1701 (2009)

\bibitem{zettl}
C.O. Girit, J. C. Meyer, R. Erni, et al. 
Science 323, 1705 (2009)

\bibitem{chuvilin}
A. Chuvilin, J. C. Meyer, G. Algara-Siller, U. Kaiser, 
New Journal of Physics 11, 083019 (2009) 



\bibitem{defects}
J. J. Palacios, J. Fern\'{a}ndez-Rossier, and L. Brey Phys. Rev. B 77, 195428 (2008).


\bibitem{esquinazi}
H. Ohldag et al., Phys. Rev. Lett. 98, 187204 (2007).

\bibitem{yazyev1} O. V. Yazyev, 
Phys. Rev. Lett. 101, 037203 (2008).

\bibitem{maggraphite}
J. S Cervenka, M. I. Katsnelson  and  C. F. J. Flipse,
Nature Physics 5, 840  (2009) 

\bibitem{graphenereviews}
A. H. Castro Neto, F. Guinea, N. M. R. Peres, K. S. Novoselov and A. K. Geim,
Rev. Mod. Phys. {\bf 81}, 109 (2009);
A. K. Geim and K. S. Novoselov et al., 
Nature Materials {\bf 6}, 183  (2007);
A. K. Geim and A. H. MacDonald,
Physics Today {\bf 60}, 35 (2007).

\bibitem{noncollinear}
K. Sawada, F. Ishii, M. Saito, S. Okada, and T. Kawai, Nano Lett. 9, 269 (2009).

\bibitem{doping}
J. Jung and A. H. MacDonald,
Phys. Rev. B 79, 235433 (2009).

\bibitem{armch_ef}
D. S. Novikov,
Phys. Rev. Lett. 99,  056802 (2007).

\bibitem{rudberg}
E. Rudberg, P. Salek, and Y. Luo,
Nano Lett., 7, 2211 (2007).

\bibitem{others_ef}
Er-Jun Kan et al. 
Appl. Phys. Lett. 91, 243116 (2007).


\bibitem{nonlocal}
J. Jung, T. Pereg-Barnea and A. H. MacDonald, to be submitted.




\end{thebibliography}
\end{document}